\documentclass[12pt]{article}
\usepackage{geometry}
\usepackage{amsmath, amssymb, amsthm, bbm}
\usepackage[dvips]{graphicx, color}
\usepackage[round]{natbib}
\usepackage{mathrsfs}
\usepackage{bm}
\usepackage[usenames, dvipsnames, svgnames, table]{xcolor}
\usepackage{mathrsfs}
\usepackage{verbatim}
\usepackage{dsfont}
\usepackage{titling}
\usepackage{txfonts}
\usepackage{threeparttable}

\usepackage{url}
\usepackage[colorlinks,
			linkcolor=blue,
			anchorcolor=blue,
			urlcolor=blue,
			citecolor=blue]{hyperref}
\usepackage{hyphenat}
\usepackage[utf8]{inputenc}
\usepackage{epstopdf}
\usepackage{booktabs} % for well-spaced horizontal rules
\usepackage{caption}
\usepackage{graphicx}
\usepackage{subfigure}
\usepackage{float}
\usepackage{listings}
\usepackage{multirow}
\usepackage{rotating}
\usepackage{longtable}
\usepackage{enumitem}
\usepackage{threeparttablex} % for "ThreePartTable" environment
\usepackage{cases}
\usepackage{filecontents}
\usepackage{algorithm}
\usepackage{algorithmic}

\usepackage{xr}
\newtheorem{theorem}{Theorem}[section]

\newtheorem{remark}{Remark}[section]

\newtheorem{assumption}{Assumption}[section]
\newtheorem{scenario}{Scenario}

\numberwithin{equation}{section}

\renewcommand{\hat}{\widehat}
\renewcommand{\tilde}{\widetilde}

\newcommand{\hDash}{\bot\!\!\!\bot}

\newcommand{\mbR}{\mathbb{R}}

\newcommand{\bX}{\mathbf{X}}

\newcommand{\bZ}{\mathbf{Z}}
\newcommand{\bV}{\mathbf{V}}
\newcommand{\bbeta}{\bm{\beta}}
\newcommand{\bgamma}{\bm{\gamma}}
\newcommand{\bSigma}{\bm{\Sigma}}

\newcommand{\mbH}{\mathbb{H}}
\newcommand{\mbE}{\mathbb{E}}
\newcommand{\mcD}{\mathcal{D}}

\newcommand{\msL}{\mathscr{L}}
\newcommand{\tr}{\mathrm{tr}}

\newcommand{\bmu}{\bm{\mu}}
\newcommand{\bA}{\mathbf{A}}
\newcommand{\bU}{\mathbf{U}}

\begin{document}
\allowdisplaybreaks[3] %allow align environment spread page

\title{\bf Tests for ultrahigh-dimensional partially linear regression models}
\author{ Hongwei Shi$^1$, Bowen Sun$^1$, Weichao Yang$^1$, and Xu Guo$^1$\thanks{Corresponding author: Xu Guo. Email address: xustat12@bnu.edu.cn.}\\
	   {\small \it $^{1}$ School of Statistics, Beijing Normal University, Beijing, China}
	   }

\date{}
\maketitle

\vspace{-0.5in}
	
\begin{abstract}
In this paper, we consider tests for ultrahigh-dimensional partially linear regression models. The presence of ultrahigh-dimensional nuisance covariates and unknown nuisance function makes the inference problem very challenging. We adopt machine learning methods to estimate the unknown nuisance function and introduce quadratic-form test statistics. Interestingly, though the machine learning methods can be very complex, under suitable conditions, we establish the asymptotic normality of our introduced test statistics under the null hypothesis and local alternative hypotheses. We further propose a power-enhanced procedure to improve the test statistics' performance. Two thresholding determination methods are provided for the power-enhanced procedure. We show that the power-enhanced procedure is powerful to detect signals under either sparse or dense alternatives and it can still control the type-I error asymptotically under the null hypothesis. Numerical studies are carried out to illustrate the empirical performance of our introduced procedures.

\medskip
		
	
\end{abstract}

\section{Introduction}
Let $Y \in \mbR$ be the response along with covariates $\bX = (X_{1}, \ldots, X_{p_1})^\top \in \mbR^{p_1}$ and $\bZ = (Z_{1}, \ldots, Z_{p_2})^\top \in \mbR^{p_2}$. We consider the following ultrahigh-dimensional partially linear model,
\begin{align}
\label{partial_LM}
Y = \bX^{\top}\bbeta + g(\bZ) + \epsilon\quad \mathrm{with} \quad \epsilon\hDash\bX.
\end{align}
Here $\bbeta\in\mathbb{R}^{p_1}$ is an unknown regression coefficient, $g(\cdot):\mbR^{p_2} \rightarrow \mbR$ is an unknown smooth function, and $\epsilon$ is a random error term with $\mbE(\epsilon)=0$ and $\mbox{Var}(\epsilon)=\sigma^2$. Due to its flexibility and interpretability, the partially linear model \citep{hardle2000partially} is widely used. In this paper, we aim to test whether the covariates of primary interest $\bX$ contribute to the response $Y$ given the other nuisance covariates $\bZ$. Under the above model (\ref{partial_LM}), this problem can be formulated as the following hypothesis testing problem,
\begin{align}
\label{H0}
\mbH_{0}:\bbeta = \bm{0},\qquad \mathrm{versus}\qquad \mbH_{1}:\bbeta \neq \bm{0}.
\end{align}
The above testing problem is of great importance in practice. For instance, in association analysis, researchers usually test the significance of a gene pathway consisting of ultrahigh-dimensional genes for the same biological function, given the other ultrahigh-dimensional genes. To address these kinds of problems, in this paper, we allow both the covariates of primary interest $\bX$ and the nuisance covariates $\bZ$ to be ultrahigh-dimensional, that is, $p_1$ and $p_2$ both can be exponential order of the sample size.

Under the high-dimensional linear or generalized linear models, many authors considered the above testing problem. Actually, \cite{zhang2017simultaneous} and \cite{dezeure2017high} introduced coordinate-based maximum tests. They first obtained debiased/desparsified Lasso estimators \citep{zhang2014confidence,van2014asymptotically,javanmard2014confidence} for each component and then used the maximum of these estimators as test statistics. This idea was also considered by many other authors. See for instance \cite{ning2017general} and \cite{ma2021global}. These maximum-type methods are computationally expensive and generally are not powerful in the presence of dense alternatives. When the alternatives are really dense, which means the parametric vector contains many non-zero small regression coefficients, the quadratic-form test procedure introduced by \cite{zhong2011tests} is a powerful approach. The procedures in \cite{zhong2011tests} are further extended by \cite{guo2016tests} to the generalized linear model. However, the procedures in \cite{zhong2011tests} and \cite{guo2016tests} can not deal with ultrahigh-dimensional nuisance covariates. To solve this critical issue, \cite{chen2022testing} introduced penalized estimators to the original procedure in \cite{guo2016tests}. Theoretically, the dimension of the parametric vector of interest can only grow polynomially with the sample size to guarantee nontrivial power.
\cite{yang2022score} made a deep theoretical investigation of the test procedure in \cite{chen2022testing}. They developed new techniques and obtained the limiting distributions under the null and local alternative hypotheses. Their results allow both $\bX$ and $\bZ$ to be ultrahigh-dimensional. However, all these papers focus on high-dimensional parametric models and can not be directly extended to the above partially linear model \eqref{partial_LM}.

There are also some studies about (\ref{H0}) under the partially linear model. See for instance \cite{wang2017generalized, wang2020test}, \cite{liu2020tests} and \cite{zhao2023new}. However, these studies only focus on low-dimensional nuisance covariates. The low-dimension nature of the nuisance covariates $\bZ$ in these papers enables the authors to estimate the unknown smooth function $g(\bZ)$ using classical nonparametric methods, such as the kernel method. However, it is widely recognized that the kernel method can fail when the dimension of $\bZ$ is relatively high. Thus these procedures can not handle the testing problem (\ref{H0}) when both the covariates of interest and the nuisance covariates are ultrahigh-dimensional.

The existing works motivate the investigation of the current paper. Different from existing papers, we allow the covariates of primary interest and the nuisance covariates to be ultrahigh-dimensional and also allow the smooth function $g(\cdot)$ to be unknown. To estimate the smooth function $g(\cdot)$ efficiently, some flexible machine learning (ML) methods are adopted. However, a direct application would make the theoretical analysis very difficult due to the complex and even black-box nature of ML methods. To this end, sample-splitting is adopted. Actually, we first randomly split the data into two parts. We use one part to estimate the smooth function $g(\cdot)$ by some ML methods. We then use the other part to construct quadratic-form test statistics. Under mild conditions, we obtain the limiting distributions of the constructed test statistics under the null and local alternative hypotheses even though we estimate the smooth function $g(\cdot)$ by some black-box ML methods.

In this paper, we further propose power-enhanced test procedures to achieve good power performance under general alternatives. Actually, when the parametric vector of interest is extremely sparse, which means only a very small subset of covariates contribute to the response, the power of quadratic-form test statistics can be low \citep{xu2016adaptive, chen2019two, yu2022power}. Although the coordinate-based maximum tests are powerful in detecting sparse alternatives, they perform poorly when the parametric vector of interest is dense. In practice, the sparsity of the parametric vector is typically unknown, making it challenging to choose a powerful test. To overcome this issue, we are inspired by the original paper of \cite{fan2015power} and introduce power-enhanced test statistics. \cite{fan2015power} considered a quadratic-form OLS-based statistic with asymptotically normal marginal distributions. However, in contrast to \cite{fan2015power}, the distributions of our marginal test statistics being degenerate $U$-statistics are no longer asymptotically normal under the null hypothesis. We need to construct power-enhanced test statistics based on degenerate $U$-statistics. For the power-enhanced procedures, one hard thresholding level and one soft thresholding level are given by investigating the tail probabilities of our marginal test statistics. We demonstrate that the enhanced test has no size distortion asymptotically, and the power is indeed enhanced under more general alternatives.

The remainder of the paper is organized as follows. In Section \ref{Construction of test statistics}, we introduce our testing~procedures and derive their limiting distributions. In Section \ref{Power-Enhanced Tests}, we propose power-enhanced tests and investigate their theoretical properties. Section \ref{Numerical Studies} presents simulation studies. Real data example is analyzed in Section \ref{Real Data Analysis}. Section \ref{Discussions and Conclusions} offers some discussions and conclusions. All the technical proofs and additional simulation results are relegated to the Supplementary Material.

\vspace{2mm}

{\noindent \bf Notations.}
The following notation is adopted throughout this paper.
For a $d$-dimensional vector $\bU = (U_1, \ldots, U_d)^{\top} \in \mbR^d$,  we define $\| \bU\|_l= (\sum_{j=1}^{d}|U_{j}|^{l})^{1/l}$ with $1 \leq l < \infty$ and $\| \bU \|_{\infty} = \max_{1\leq j\leq d}| U_{j}|$ to denote $L_q$ and $L_{\infty}$ norms of $\bU$.
A random variable $X$ is \textit{sub-Gaussian} if the moment generating function (MGF) of $X^2$ is bounded at some point, namely $\mathbb{E}\exp(X^2/K^2)\leq2$, where $K$ is a positive constant. A random variable $Y$ is $\operatorname{\mathit{sub-Exponential}}$ if the MGF of $\lvert Y\rvert$ is bounded at some point, namely $\mathbb{E}\exp(\lvert Y\rvert/K')\leq2$, where $K'$ is a positive constant. The \textit{sub-Gaussian norm} of a \textit{sub-Gaussian} random variable $X$ is defined as $\|X\|_{\psi_2} = \inf\{t>0:\mbE\exp(X^2/t^2)\leq2\}$.
For $q_1\times q_2$-dimensional matrix $\bA$, we define $\|\bA\|_{\infty} = \max_{ij}|A_{ij}|$, where $A_{ij}$ is the $(i,j)$-th element of $\bA$.
For notational simplicity, we let $c$ and $C$ be generic constants.

\section{Quadratic-form test statistics and their properties}
\label{Construction of test statistics}

Suppose that $\mcD=(\bX_i,\bZ_i,Y_i)_{i=1}^N$ with $N=2n$ is an independent and identically distributed random sample from the population $(\bX,\bZ,Y)$. Denote $S(\bZ) = \{Y - g(\bZ)\}\bX$. Under the null hypothesis, $\mbE[S(\bZ)]^{\top}\mbE[S(\bZ)] = 0$, while under the alternative hypothesis, $\mbE[S(\bZ)]^{\top}\mbE[S(\bZ)] > 0$. This then implies that we can construct test statistics based on sample versions of $\mbE[S(\bZ)]^{\top}\mbE[S(\bZ)]$. Inspired by  \cite{zhong2011tests}, \cite{guo2016tests}, \cite{chen2022testing} and \cite{yang2022score}, we may consider the following quadratic-form test statistic,
\begin{align}
\label{TN}
T_N = \frac{1}{N}\sum_{i\neq j}\{Y_i - \hat{g}(\bZ_i)\}\{Y_j - \hat{g}(\bZ_j)\}\bX_i^\top\bX_j.
\end{align}
Here $\hat{g}(\cdot)$ is a suitable estimator of $g(\cdot)$ obtained from $(\bZ_i, Y_i)_{i=1}^N$. {The advantage of not using $\bX_i, i=1,\cdots,N$ for estimation is that we can sufficiently exploit the model structure under the null hypothesis and avoid imposing strong assumptions on the dimension and sparsity of $\bbeta$.}

Since the dimension of $\bZ$ is very high, classical nonparametric methods such as the kernel method fail. To handle this issue, we use some flexible machine learning (ML) methods such as Lasso, Random Forest, and Neural Nets to obtain the estimator $\hat{g}(\cdot)$. However, a direct involvement of $\hat{g}(\cdot)$ in the above $T_N$ makes it very difficult to analyze the properties of the test statistic $T_N$ due to the possible over-fitting/high-complexity phenomena that commonly occur in ML estimates. To this end, the sample-splitting idea is adopted to separate the estimation of the nuisance function from the construction of test statistics. The sample-splitting idea is a powerful approach that has been employed by many authors in different problems. See for instance \cite{fan2012variance}, \cite{chernozhukov2018double}, \cite{du2021false} and \cite{cai2023tests}.

To be precise, we first randomly split data $\mcD$ into two parts: $\mcD_1$ and $\mcD_2$ with equal size, $|\mcD_k|=n, k=1,2$. We estimate $g(\cdot)$ by some ML algorithms based on the data $\mcD_k$ and denote the estimator as $\hat{g}_k(\cdot)$. We then construct test statistics based on the other part $\mcD_{3-k}$. That is, we consider the following test statistics,
\begin{align}
\label{Tnk}
T_{nk} = \frac{1}{n}\sum_{i\neq j, i, j \in \mcD_{3-k}}\{Y_i - \hat{g}_k(\bZ_i)\}\{Y_j - \hat{g}_k(\bZ_j)\}\bX_i^\top\bX_j, \quad k = 1, 2.
\end{align}

The above test statistics $T_{nk}$ are also inspired by the elegant double machine learning framework in \cite{chernozhukov2018double}. In \cite{chernozhukov2018double}, the dimension of interested covariates $\bX$ is fixed. While here we focus on the situation that $\bX$ is ultrahigh-dimensional. One may extend the approach of \cite{chernozhukov2018double} by adopting the coordinate-based maximum tests. That is, for each element of $\bX$, use the double machine learning approach to construct element-wise test statistics and then consider the maximum of these element-wise test statistics. Compared with coordinate-based maximum tests, here we only need to estimate $g(\cdot)$ twice, and thus the computational cost is significantly reduced. Further as discussed in the Introduction, coordinate-based maximum tests are not powerful for dense alternatives.

In the following, we will show that the above constructed test statistics $T_{nk}$ are asymptotically normal under the null hypothesis even if we estimate the smooth function $g(\cdot)$ by black-box ML techniques.

\subsection{Asymptotic null distribution}

Let $\bV = (\bX^\top,\bZ^\top)^\top$, $\bSigma_{\bX} = \mbE(\bX\bX^\top)$, $\bSigma_{\bX \bZ} = \mbE(\bX\bZ^\top)$, $\bSigma_{\bZ \bX} = \mbE(\bZ\bX^\top)$ and $p =p_1+p_2$. Denote $\bmu_{gk} = \mbE[\{g(\bZ) - \hat{g}_k(\bZ)\}\bX]$ and $\Lambda_{\bX} = 2\sigma^4\tr(\bSigma_{\bX}^2)$. We impose the following technical assumptions to study the asymptotic null distributions of the test statistics $T_{nk} \ (k = 1, 2)$.

\begin{assumption}
\label{assumption1}
Assume that $\tr(\bSigma^4_{\bX})=o(\tr^2(\bSigma_{\bX}^2))$ and $\tr(\bSigma_{\bX}^2)\rightarrow\infty$ as $(n,p_1)\rightarrow\infty$.
\end{assumption}

\begin{assumption}
\label{assumption2}
Let $\bA_{1}$ and $\bA_2$ be two $p_1\times p_1$ semi-positive matrices. Assume that $\mbE\biggl(\prod_{i=1}^{2}\bX^\top\bA_{i}\bX\biggr) \leq C\prod_{i=1}^{2}\tr(\bA_{i}\bSigma_{\bX})$.
\end{assumption}

\begin{assumption}
\label{assumption3}
Assume that $n\bmu_{gk}^{\top}\bmu_{gk} = o(\sqrt{\Lambda_{\bX}})$.
\end{assumption}

\begin{assumption}
\label{assumption4}
Assume that $c<\mbE(\epsilon^2)$ and $\mbE(\epsilon^4)<C$ for some positive constants $c, C$.
\end{assumption}

\begin{assumption}
\label{assumption5}
Assume that $\mbE[\{g(\bZ) - \hat{g}_k(\bZ)\}^4] = o(1)$.
\end{assumption}

Assumption \ref{assumption1} is a mild condition that is commonly imposed in the literature \citep{zhong2011tests, guo2016tests, cui2018test}. Note that if all the eigenvalues of $\bSigma_{\bX}$ are bounded, then $\tr(\bSigma^4_{\bX})=o(\tr^2(\bSigma_{\bX}^2))$ holds trivially.
In Assumption \ref{assumption2}, we only make a very mild moment condition and make no distribution assumption on $\bX$ nor $\bZ$. Note that this assumption is weaker than the pseudo-independence assumption and the elliptical distribution assumption, which are commonly assumed in the literature \citep{zhong2011tests, guo2016tests, cui2018test}.
Assumption \ref{assumption3} is imposed to control the impact of the bias from $\hat{g}_k(\bZ)$ on the asymptotic distributions of $T_{nk}$. In the following Remark \ref{remark1}, we make further discussions about this assumption. In
Assumption \ref{assumption4}, mild moment conditions for the error term are imposed.
Assumption \ref{assumption5} is a mild condition on the rate of estimating the nuisance function $g(\cdot)$, which is available for many ML methods.
In the above assumptions, the covariates of primary interest $\bX$ and the nuisance covariates $\bZ$ are both allowed to be ultrahigh-dimensional.

\begin{remark}
\label{remark1}
Suppose that the model \eqref{partial_LM} is linear with $g(\bZ)=\bZ^\top\bgamma$, where $\bgamma$ is a $p_{2}$-dimensional regression coefficient. Further let $\hat\bgamma_k$ be a penalized estimator of $\bgamma$. By H{\"o}lder's inequality, we have
\begin{align*}
n\bmu_{gk}^\top\bmu_{gk}=n(\bgamma-\hat{\bgamma}_k)^\top
\mbE(\bZ\bX^\top)\mbE(\bX\bZ^\top)(\bgamma-\hat{\bgamma}_k)
\leq n\|\bgamma-\hat\bgamma_k\|_1^2\|\bSigma_{\bZ\bX}\bSigma_{\bX\bZ}\|_{\infty}.
\end{align*}
Note that
\begin{align*}
(\bSigma_{\bZ\bX}\bSigma_{\bX\bZ})_{ij} =\sum_{l}\mbE(Z_iX_l)\mbE(X_l Z_j)
\leq \frac{1}{2}\sum_l\mbE^2(Z_iX_l)+\mbE^2(X_lZ_j)
\leq \varrho^2.
\end{align*}
Here $\varrho^2 = \max_{1\leq h\leq p_{2}}\|\mbE(Z_{h}\bX)\|_2^2$. Generally, it holds that $\|\bgamma-\hat\bgamma_k\|_1=O_p(s_{2}\sqrt{\log p_{2}/n})$, where $s_2$ is the sparsity level of $\bgamma$. Many penalized estimators, such as Lasso, SCAD and MCP, can achieve this $L_1$ error bound of $\hat\bgamma_k$. See for instance \cite{fan2020statistical}. A sufficient condition for Assumption \ref{assumption3} thereby is $s_{2}^2\log p_{2}\varrho^2=o(\sqrt{\Lambda_{\bX}})$. When the eigenvalues of the variance matrix of $\bV$ are all bounded, $\varrho^2$ is bounded and $\Lambda_{\bX}$ is in the order of $p_1$. Then the condition $s_{2}^2\log p_{2}\varrho^2=o(\sqrt{\Lambda_{\bX}})$ can be simplified as $s^2_2\log p_{2} =o(\sqrt{p_1}).$  When $p_1\geq c n^2$, this condition is then even weaker than $s_2^2\log p_2=o(n)$. In other words, Assumption \ref{assumption3} holds when the covariates $\bX$ and $\bZ$ are both ultrahigh-dimensional, reaching the exponential order of the sample size.
\end{remark}

\begin{theorem}
\label{thm1}
Under $\mbH_0$, if Assumptions \ref{assumption1}-\ref{assumption5} hold, we have
\begin{align}
\label{null_asym}
\frac{T_{nk}}{\sqrt{\Lambda_{\bX}}} \stackrel{d}{\rightarrow} N(0,1) \quad \text {as} \  (n, p_1, p_2) \rightarrow \infty,
\end{align}
where $T_{nk}$ is defined in \eqref{Tnk} and $\Lambda_{\bX} = 2\sigma^4\tr(\bSigma_{\bX}^2)$.
\end{theorem}

In the above Theorem, we establish the limiting null distribution of the proposed test statistics $T_{nk} \ (k = 1, 2)$. Notably, our test statistics allow very flexible and complex ML methods.

For inference, we need an estimate of $\Lambda_{\bX}$ involved in Theorem \ref{thm1}. We define
\begin{align}
\label{Lambda_k_hat}
\hat{\Lambda}_{\bX, k} = 2\hat{\sigma}_k^4\hat{\tr(\bSigma_{\bX}^2)}_k.
\end{align}
Here we use the ratio consistent estimator of $\tr(\bSigma^2_{\bX})$ following \cite{zhong2011tests}, that is
\begin{align*}
\hat{\tr(\bSigma^2_{\bX})}_k=\frac{1}{2 \binom{n}{4}}\sum_{i_1< i_2< i_3< i_4 \in \mcD_{3-k}}
(\bX_{i_1}-\bX_{i_2})^\top(\bX_{i_3}-\bX_{i_4})(\bX_{i_2}-\bX_{i_3})^\top(\bX_{i_4}-\bX_{i_1}).
\end{align*}
While for the variance $\sigma^2$, we can estimate it as follows,
\begin{align*}
\hat\sigma^2_k=\frac{1}{n}\sum_{i\in\mcD_{3-k}}\{Y_i-\hat{g}_k(\bZ_i)\}^2.
\end{align*}
Next we obtain the normalized test statistics $\tilde{T}_{nk} \ (k=1, 2)$ by plugging the estimator of $\Lambda_{\bX}$ into \eqref{null_asym}. From Theorem \ref{thm1} and the Slutsky Theorem, we further conclude that
\begin{align}
\label{Tnk_tilde}
\tilde{T}_{nk} = \frac{T_{nk}}{\sqrt{\hat{\Lambda}_{\bX, k}}} \stackrel{d}{\rightarrow} N(0, 1) \quad \text {as} \  (n, p_{1}, p_{2}) \rightarrow \infty.
\end{align}

It is noteworthy that the data-splitting statistics given above have the drawback that only the information from half of the data is exploited to make inference, which can lead to a significant power loss. To deal with this issue, we can utilize cross-fitting to achieve further improvement. See for instance \cite{fan2012variance}, \cite{chernozhukov2018double} and \cite{cai2023tests}. As above, we can construct two test statistics $\tilde{T}_{n1}$ and $\tilde{T}_{n2}$. Further note that they are asymptotically independent. From Theorem \ref{thm1} and by averaging the two resulting statistics, we have the following limiting distribution,
\begin{align}
\label{Tn_tilde}
\tilde{T}_{n} = \frac{1}{\sqrt{2}} (\tilde{T}_{n1} + \tilde{T}_{n2}) \stackrel{d}{\rightarrow} N(0, 1), \quad \text {as} \  (n, p_{1}, p_{2}) \rightarrow \infty.
\end{align}
Hence an asymptotic $\alpha$-level test rejects $\mbH_0$ if
\begin{align*}
\tilde{T}_{n}\geq z_{\alpha}.
\end{align*}
Here $z_{\alpha}$ is the upper-$\alpha$ quantile of the standard normal distribution, and $\alpha$ is the significance level. For the sake of clarity, the detailed algorithm for making inference on $\bbeta$ is provided in Algorithm \ref{alg1}.

\begin{algorithm}[htb]
\renewcommand{\algorithmicrequire}{\textbf{Input:}}
	\renewcommand{\algorithmicensure}{\textbf{Output:}}
\caption{Inference procedure for $\mbH_0$}
\begin{algorithmic}
\label{alg1}
\REQUIRE Data $\mcD = (\bX_i,\bZ_i,Y_i)_{i=1}^{N}$ with $N=2n$.
\ENSURE $p$-value for $\mbH_{0}:\bbeta = \bm{0}$.

\STATE \textbf{Step 1.} Randomly split the data $\mcD$ into two parts $\mcD_1$ and $\mcD_2$ with equal sample size $n$.
\STATE \textbf{Step 2.} Estimate the nuisance function $g(\cdot)$ by some ML algorithms on the data $\mcD_k$, and denote~the estimator as $\hat{g}_k(\cdot)$.
\STATE \textbf{Step 3.} Construct the test statistic $T_{nk}$ as shown in \eqref{Tnk} based on the data $\mcD_{3-k}$. Calculate the test statistic $\tilde{T}_{nk}$ using the formulas \eqref{Lambda_k_hat} and \eqref{Tnk_tilde}.
\STATE \textbf{Step 4.} Let $k=1, 2$ and repeat \textbf{Step 2} and \textbf{Step 3} to obtain the test statistics $\tilde{T}_{n1}$ and $\tilde{T}_{n2}$. Further calculate $\tilde{T}_{n} = (\tilde{T}_{n1} + \tilde{T}_{n2})/\sqrt{2}$.
\STATE \textbf{Step 5.} Calculate the $p$-value by $1-\Phi(\tilde{T}_{n})$, where $\Phi(\cdot)$ is the cumulative distribution function of the standard normal distribution. The null hypothesis $\mbH_0$ is rejected if $1-\Phi(\tilde{T}_{n}) \leq \alpha$.
\end{algorithmic}
\end{algorithm}

To reduce the randomness induced by data splitting, we further consider an ensemble testing procedure based on multiple data splitting. This is an adoption of the procedure in \cite{meinshausen2009p}. The detailed algorithm is presented in the Supplementary Material. With this ensemble testing procedure, the type-I error can still be asymptotically controlled.

\subsection{Power analysis}
In this subsection, we present power analysis for the proposed test statistics. Consider the following family of local alternatives,
\begin{align*}
\msL(\bbeta) = \left\{\bbeta\in\mbR^{p_{1}}\bigg| \bbeta^\top\bSigma_{\bX}\bbeta = o(1), \ \bbeta^\top\bSigma_{\bX}^{2}\bbeta = o\left(\frac{\Lambda_{\bX}}{n^2\bmu_{gk}^{\top}\bmu_{gk}}\right)\  \text{and} \ \bbeta^\top\bSigma_{\bX}^3\bbeta = o\left(\frac{\Lambda_{\bX}}{n}\right)\right\}.
\end{align*}
The following theorem gives the asymptotic behaviour of the proposed test statistics $T_{nk} \ (k=1, 2)$ under the above local alternative hypotheses.

\begin{theorem}
\label{thm2}
Under conditions in Theorem \ref{thm1}, and for $\bbeta\in\mathscr{L}(\bbeta)$, we have
\begin{align*}
\frac{T_{nk}- n\bbeta^\top\bSigma_{\bX}^{2}\bbeta}{\sqrt{\Lambda_{\bX}}}\stackrel{d}{\rightarrow} N(0,1) \quad \text {as} \  (n, p_{1}, p_{2}) \rightarrow \infty.
\end{align*}
\end{theorem}

Regard $n\bbeta^{\top}\bSigma_{\bX}^2\bbeta$ as the signal strength and the power is largely determined by the following signal-to-noise ratio,
\begin{align*}
\mathrm{SNR} = \frac{n\bbeta^{\top}\bSigma_{\bX}^2\bbeta}{\sqrt{\Lambda_{\bX}}}.
\end{align*}
Thus when the signal-to-noise ratio is large, our proposed test statistics can have large powers for the alternative hypotheses $\mathscr{L}(\bbeta)$.

In above theorems, we assume $n\bmu_{gk}^\top\bmu_{gk}=o(\sqrt{\Lambda_{\bX}})$ in Assumption \ref{assumption3}. However, this assumption may not hold under certain alternative hypotheses. In the following, we investigate the performance of $T_{nk}$ when we can not estimate $g(\cdot)$ very well.

\begin{assumption}
\label{assumption6}
Assume that $\mbE[\{g(\bZ) - \hat{g}_k(\bZ)\}^4] = O(1)$.
\end{assumption}
In Assumption \ref{assumption5}, we require that $\mbE[\{g(\bZ) - \hat{g}_k(\bZ)\}^4] = o(1)$. But in above Assumption \ref{assumption6}, we allow $\hat{g}_k(\bZ)$ estimate $g(\bZ)$ even inconsistently. Actually since we obtain $\hat g_k(\bZ)$ from $(\bZ_i, Y_i), i\in\mcD_k$, $\hat g_k(\bZ)$ aims to estimate $g^*(\bZ)=\mbE(Y|\bZ)=g(\bZ)+\mbE(\bX|\bZ)^\top\bbeta$, which is different from $g(\bZ)$ under alternative hypothesis.

We consider the following family of alternative hypotheses,
\begin{align*}
\msL^{\text{A}}(\bbeta) = \left\{\bbeta\in\mbR^{p_{1}}\Big| \sqrt{\Lambda_{\bX}}=o\big(n\|\bSigma_{\bX}\bbeta+\bmu_{gk}\|_2^2\big), \bbeta^{\top}\bSigma_{\bX} \bbeta = O(1) \right\}.
\end{align*}
The above alternative hypotheses correspond to the situation that either the signal-to-noise ratio,
${n\bbeta^{\top}\bSigma_{\bX}^2\bbeta}/{\sqrt{\Lambda_{\bX}}}$ is very large or the bias term $n\bmu_{gk}^{\top}\bmu_{gk}$ from $\hat{g}_k(\cdot)$ relative to $\sqrt{\Lambda_{\bX}}$ is large. The following theorem shows that under the above alternatives, the power of our proposed test statistics can be asymptotically 1.

\begin{theorem}
\label{thm3}
If Assumptions \ref{assumption2}, \ref{assumption4} and \ref{assumption6} hold, and for $\bbeta \in \msL^{\text{A}}(\bbeta)$, we have
\begin{align*}
\frac{T_{nk}}{\sqrt{\Lambda_{\bX}}}{\rightarrow} \infty \quad \text {as} \  (n, p_{1}, p_{2}) \rightarrow \infty.
\end{align*}
\end{theorem}

This theorem implies that even if we can not estimate the function $g(\cdot)$ well under alternative hypotheses, we can still have high detection power. Further note that in the above theorem, Assumption \ref{assumption1} is not required. From the theoretical analysis, it is known that Assumption \ref{assumption1} is adopted to establish the asymptotic normality of our proposed test statistics by applying the martingale central limit theorems, while in the above theorem, we only need to show that the variance of $T_{nk}$ is dominated by the expectation of $T_{nk}$.

\section{Power-enhanced tests}
\label{Power-Enhanced Tests}

This section discusses power-enhanced tests and presents the theoretical properties of power-enhanced statistics. Two techniques to determine the threshold are also provided.

In Section \ref{Construction of test statistics}, we establish some promising asymptotic properties for the quadratic-form test statistics $T_{nk}$, such quadratic-form statistics work powerfully against dense alternatives. However, they generally perform poorly against sparse alternatives. For relevant discussions, see \cite{xu2016adaptive}, \cite{chen2019two} and \cite{yu2022power}. One typical approach to achieve high detection power against sparse alternatives is to use maximum-form statistics. However, as discussed in the Introduction, coordinate-based maximum tests generally have heavy computation burdens. Further, in practice, the sparsity of alternative hypotheses is unknown. We aim to improve the power performance of the quadratic-form test statistics $T_{nk}$ for sparse alternative hypotheses.

Recently, \cite{fan2015power} introduced a very elegant power enhancement procedure. Actually, they achieved power enhancement by adding a constructed component to an asymptotically pivotal statistic. As long as the constructed component is positive, the power is boosted. However, a reasonable component should have no size distortion. To achieve this goal, the construction of the power enhancement component relies on screening over the marginal test statistics. In \cite{fan2015power}, their marginal test statistics are asymptotically normal. Nevertheless, in our paper, the marginal test statistics are degenerate $U$-statistics, and thus the distributions of our marginal test statistics are no longer asymptotically normal under the null hypothesis. We need more efforts in the design of power enhancement components based on degenerate $U$-statistics.

Now we define the marginal test statistic as
\begin{align}
\label{Tlk}
T_{lk}= \frac{1}{n(n-1)}\sum_{i\neq j, i, j \in \mcD_{3-k}}\{Y_i - \hat{g}_k(\bZ_i)\}\{Y_j - \hat{g}_k(\bZ_j)\}X_{il}X_{jl},
\end{align}
for $k=1, 2$ and $l = 1, \ldots, p_1$. Here $X_{il}$ is the $l$-th component of $\bX_i$. Further, let
\begin{align}
\label{T0k}
T_{0k}=a_{n,p}\sum_{l=1}^{p_1}|T_{lk}| I(|T_{lk}| > \delta_k) \geq 0,
\end{align}
where $\delta_k > 0$ is a critical thresholding level to strike a balance between removing non-signal $T_{lk}$'s and keeping those with signals. While $a_{n,p} > 0$ is a user-specific value. The determination of the threshold $\delta_k$ is discussed in the next subsection. The choice of such threshold is straightforward for test statistics which asymptotically follow well-known normal distributions but requires additional efforts for degenerate $U$-statistics which follow non-normal distributions. The determination of threshold relies on the tail probabilities of marginal test statistics $T_{lk}$'s. We control the tail probabilities of $T_{lk}$ by the Hanson-Wright inequality to quadratic forms in random variables with $\alpha$-sub-exponential tail decay \citep{gotze2021concentration}. The techniques developed in this paper can be useful in other inference problems.

The power-enhanced test statistic is then defined as
\begin{align}
\label{TPEk}
T_{\mathrm{PE},k}=\tilde{T}_{nk} + T_{0k} = \frac{T_{nk}}{\sqrt{2\hat{\sigma}_k^4\hat{\tr(\bSigma_{\bX}^2)}_k}} + T_{0k}, \quad k = 1, 2.
\end{align}
Here the non-negativeness of $T_{0k}$ ensures that $T_{\mathrm{PE},k}$ is at least as powerful as $\tilde{T}_{nk}$.
Same to \eqref{Tn_tilde}, we can also adopt the idea of cross-fitting and further construct
\begin{align}
\label{TPE}
T_{\mathrm{PE}} = \frac{1}{\sqrt{2}}(T_{\mathrm{PE},1} + T_{\mathrm{PE},2}).
\end{align}

\subsection{Theoretical properties}

We denote $\mu_{gk, l}=\mbE[\{g(\bZ)-\hat{g}_k(\bZ)\} X_{l}]$ and present the following assumptions to establish the theoretical properties regarding the power enhancement component $T_{0k}$.

\begin{assumption}
\label{assumptionp1}
Assume that $\epsilon_i$, $X_{il}$'s and given $\mathcal{D}_{k}$, $g(\bZ_i)-\hat{g}_k(\bZ_i)$ are all sub-Gaussian with uniformly bounded sub-Gaussian norm.
\end{assumption}

\begin{assumption}
\label{assumptionp2}
Assume that $\max_l \mu_{gk,l}^2=o(\delta_k)$.
\end{assumption}

The sub-Gaussian assumption in Assumption \ref{assumptionp1} is imposed to control the tail probabilities of marginal statistics, ensuring the power enhancement component $T_{0k}$ equal to zero asymptotically under the null hypothesis. This sub-Gaussian assumption is standard in high-dimensional analysis \citep{wainwright2019high}. By Cauchy-Schwarz inequality, $\max_l\mu_{gk, l}^2\leq \mbE[\{g(\bZ)-\hat{g}_k(\bZ)\}^2]\max_l\mbE[ X^2_{l}]\leq C\mbE[\{g(\bZ)-\hat{g}_k(\bZ)\}^2]$. From \cite{chernozhukov2018double} and \cite{vansteelandt2022assumption}, $\mbE[\{g(\bZ)-\hat{g}_k(\bZ)\}^2]=o(n^{-1/2})$ for many ML methods. While from the tail probabilities of $T_{lk}$, we can set $\delta_k$ in the order of $(\log p_1)^2/n$. Thus when $\log p_1$ is larger than the order of $n^{1/4}$, this assumption holds. Further under linear model $g(\bZ)=\bZ^\top\bgamma$, $\mbE[\{g(\bZ)-\hat{g}_k(\bZ)\}^2]=o(s_2\log p_2/n)$. Then a sufficient condition for Assumption \ref{assumptionp2} is $s_2\log p_2=o((\log p_1)^2)$. When $p_1$ is larger than $p_2$ and the sparsity level $s_2$ is relatively small, Assumption \ref{assumptionp2} also holds.

In the following theoretical properties of $T_{0k}$, we set $\delta_k = {(\log p_1)^2}/{n}$, which is determined by the tail probability of marginal test statistics $T_{lk}$'s.

\begin{theorem}
\label{thm3.1}
Under Assumptions \ref{assumptionp1} and \ref{assumptionp2}, we have
\begin{align*}
\Pr(T_{0k}=0|\mbH_0)\rightarrow 1.
\end{align*}
\end{theorem}

Theorem \ref{thm3.1} proves that $T_{0k}=0$ holds under $\mbH_0$ with
probability tending to 1. Thus, adding $T_{0k}$ to the test statistic $\tilde{T}_{nk}$ will not affect its limiting null distribution. The
proposed power-enhanced test statistic rejects $\mbH_0$ with the significance level $\alpha$ if $T_{\mathrm{PE}} >z_{\alpha}$.

To establish the power enhancement property for $T_{\mathrm{PE},k}$, we impose the following moment assumption.
\begin{assumption}
\label{assumptionp3}
Assume that $\mbE[\{Y-\hat g_k(\bZ)\}^2\bX^\top\bX]\leq C\tr(\bSigma_{\bX})$.
\end{assumption}
Clearly from Cauchy-Schwarz inequality, a sufficient condition for this assumption is that $\mbE\{Y-\hat g_k(\bZ)\}^4<C$ and $\mbE[(\bX^\top\bX)^2]\leq C\tr^2(\bSigma_{\bX})$. In what follows, we consider the parameter spaces $\msL_s(\bbeta)$ for the sparse local alternatives,
\begin{align*}
\msL_s(\bbeta) = \left\{\bbeta\in\mbR^{p_1}\bigg\vert \Delta^2>64\delta_k\right\},
\end{align*}
where $\Delta=\max_l |\mu_{gk,l}+\mbE(X_l\bX)^{\top}\bbeta|$, $l = 1, \ldots, p_1$.

\begin{theorem}
\label{thm3.2}
Under Assumption \ref{assumptionp1} and \ref{assumptionp3}, with $a_{n,p}\delta_k/\sqrt{p_1}\rightarrow \infty$, we have
\begin{align*}
\inf_{\bbeta\in \msL_s(\bbeta)}\Pr(T_{\mathrm{PE},k}>z_{\alpha})\rightarrow 1.
\end{align*}
\end{theorem}

Since $T_{0k}\geq 0$, we always have $\Pr(T_{\mathrm{PE},k}>z_{\alpha}|\mbH_1)
\geq \Pr(\widetilde{T}_{nk}>z_{\alpha}|\mbH_1)$. That is, $T_{\mathrm{PE},k}$ is always more powerful than $\widetilde{T}_{nk}$ for any alternative hypothesis. Further, the power-enhanced test $T_{\mathrm{PE},k}$ is also powerful to detect sparse alternative hypotheses in $\msL_s(\bbeta)$.

We make some discussions about the alternative hypotheses in $\msL_s(\bbeta)$. The $\Delta$ is the largest signal brought from our marginal test statistics $T_{lk}$'s.
If $\max_l \mu_{gk,l}^2$ is obviously larger than $\delta_k=(\log p_1)^2/n$, the power-enhanced test $T_{\mathrm{PE},k}$ can reject the null hypothesis. This can occur under alternative hypotheses. On the other hand, if Assumption \ref{assumptionp2} still holds, $\Delta$ is determined by $\max_l |\mbE(X_l\bX)^\top\bbeta|$. Consider an extremely sparse situation, that is, $\bbeta=(\beta_1,0,\cdots,0)^\top$. We then have
$\max_l |\mbE(X_l\bX)^\top\bbeta|=\max_l|\mbE(X_l X_1)\beta_1|
\geq \mbE(X_1^2)\beta_1$. Thus in this toy example, as long as $\mbE(X_1^2)\beta_1$ is larger than $64\delta_k$, we can detect the alternative.

\subsection{Determination of threshold}

It is important to note that the power of $T_{\mathrm{PE},k}$ is enhanced without inflating the size asymptotically. Actually, the key to boosting the power while controlling the size asymptotically is to select a proper threshold $\delta_k$ in \eqref{T0k}. In practice, a small threshold may fail to control the type-I error, whereas a large threshold would cause no power enhancement at all. In this subsection, we introduce two innovative methods for determining the threshold: \textit{hard threshold} and \textit{soft threshold}.

At first, we use a thresholding level $\delta_k^\mathrm{hard}=\lambda_k\log\log n\frac{(\log p_1)^2}{n}$ where $\lambda_k \in (0,1]$ and a slightly larger $\log\log n$ is multiplied for the purpose of mitigating finite-sample biases \citep{fan2015power}, namely \textit{hard threshold}. This choice is simple to implement in practice but depends on user-specific $\lambda_k$. To address this limitation, we further suggest a data-driven technique to determine the threshold, namely \textit{soft threshold}, which can accommodate different data structures. Actually, we provide a bootstrap calibration for the marginal statistic $T_{lk}$ as shown in \eqref{Tlk} under $\mbH_0$. Now denote
\begin{align}
\label{Tlk_star}
T_{lk}^{*}= \frac{1}{n(n-1)}\sum_{i\neq j, i, j \in \mcD_{3-k}}\{Y_i - \hat{g}_k(\bZ_i)\}\{Y_j - \hat{g}_k(\bZ_j)\}X_{il}X_{jl}e^{*}_ie^{*}_j
\end{align}
for $k=1, 2$ and $l = 1, \ldots, p_1$. Here $(e_{i}^{*})_{i=1}^{n}$ is a sequence of independent standard normal random variables that is independent of $(\bX_{i}, \bZ_{i}, Y_{i})_{i=1}^{N}$. We then compute the maximum among the bootstrap calibrations of all marginal statistics as
\begin{align*}
\delta_k^{*} = \max_{1\leq l \leq p_1} |T_{lk}^{*}|, \quad k = 1, 2.
\end{align*}
Next, we repeat many times to obtain lots of $\delta_k^{*}$, denoted as $\{\delta_{k(1)}^{*}, \delta_{k(2)}^{*}, \ldots, \delta_{k(R)}^{*}\}$ where $R$ is the number of replicates. Finally, we set the soft threshold as
\begin{align*}
\delta_k^{\mathrm{soft}} = \max_{1\leq r \leq R}\delta_{k(r)}^{*}.
\end{align*}
As a rule of thumb, we suggest $R=30$. The detailed procedure is described in Algorithm \ref{alg2}.

\begin{algorithm}[htb]
\renewcommand{\algorithmicrequire}{\textbf{Input:}}
	\renewcommand{\algorithmicensure}{\textbf{Output:}}
\caption{The determination of soft threshold $\delta_k^{\mathrm{soft}}$}
\begin{algorithmic}
\label{alg2}
\REQUIRE Data $\mcD = (\bX_i,\bZ_i,Y_i)_{i=1}^{N}$ with $N=2n$.
\ENSURE The soft threshold $\delta_k^{\mathrm{soft}}$ in \eqref{T0k}.

\STATE \textbf{Step 1.} Randomly split the data $\mcD$ into two parts $\mcD_1$ and $\mcD_2$ with equal sample size $n$.
\STATE \textbf{Step 2.} Estimate the nuisance function $g(\cdot)$ by some ML methods on the data $\mcD_k$, and denote~the estimator as $\hat{g}_k(\cdot)$.
\STATE \textbf{Step 3.} Generate a vector of $n$ random errors $(e_i^{*})_{i=1}^n$ from the standard normal distribution. Then construct the bootstrap calibration $T_{lk}^{*}$ based on the data $\mcD_{3-k}$ for each covariate $X_l$ with $l=1, \ldots, p_1$. See \eqref{Tlk_star} for the formula.

\STATE \textbf{Step 4.} Compute the maximum of $T_{lk}^{*}$ with $l=1, \ldots, p_1$, denoted as $\delta_k^{*} = \max_{1\leq l \leq p_1} |T_{lk}^{*}|$.
\STATE \textbf{Step 5.} Repeat \textbf{Step 4} $R$ times to obtain a set of $\delta_k^{*}$, denoted as $\{\delta_{k(1)}^{*}, \delta_{k(2)}^{*}, \ldots, \delta_{k(R)}^{*}\}$. Then calculate the maximum of all $\delta_{k(\cdot)}^{*}$ to derive the soft threshold as $\delta_k^{\mathrm{soft}} = \max_{1\leq r \leq R}\delta_{k(r)}^{*}$.
\end{algorithmic}
\end{algorithm}

\section{Numerical studies}
\label{Numerical Studies}

In this section, we present some simulation experiments to illustrate the performance of our proposed inference procedures. To implement the methods in our article, R code can be downloaded in \url{https://github.com/havanashw/TestPLM}.

\subsection{Basic setups}
Throughout the simulation study, we generate the data from the model \eqref{partial_LM},
\begin{align*}
Y_i = \bbeta^\top\bX_i + g(\bZ_i) + \epsilon_i,
\end{align*}
where the predictors $(\bX_i \in \mbR^{p_1}, \bZ_i \in \mbR^{p_2})$ with $p_1 + p_2 = p$ are simulated from the multivariate normal distribution $N_p(\bm{0}_p, \bSigma)$. Here the covariance matrix $\bSigma$ is a $p \times p$-dimensional covariance matrix following the Toeplitz structure, $(\bSigma)_{ij}=\rho^{|i-j|}$ for $i,j = 1, \ldots, p$. Furthermore, the~random error term $\epsilon_i$ being independent with $(\bX_i, \bZ_i)$, is generated according to the standard normal distribution $N(0,1)$. For considerations of $g(\bZ_i)$, the following three forms are set up,
\begin{align*}
&\textbf{Model 1: } g(\bZ_i)=\bZ_i^\top \bgamma/3;\\
&\textbf{Model 2: } g(\bZ_i)=\cos(\bZ_i^\top \bgamma/2)\log(|\bZ_i^\top \bgamma| + 1);\\
&\textbf{Model 3: } g(\bZ_i)=\frac{Z_{i1}+Z_{i2}+Z_{i3}}{1+\exp(Z_{i4}+Z_{i5}+Z_{i6})}.
\end{align*}
Here $Z_{ij}$ is the $j$-th element of $\bZ_i$. Note that Model 1 corresponds to a linear model, while Models 2 and 3 are nonlinear models. The coefficients of interested predictors $\bbeta$ are generated from $\beta_{j}=c_1$ for $1 \leq j \leq s_{1}$ and $\beta_{j}=0$ otherwise. Without loss of generality, we also generate the coefficients of nuisance parameters $\bgamma$ in Models 1 and 2 from the same structure as $\bbeta$, that is $\bgamma$ has $s_{2}$ equal nonzero elements, denoted as $c_2$.
Denote $s_{1}$ and $s_{2}$ be the sparsity levels of $\bbeta$ and $\bgamma$, respectively.

Our goal is to test the hypothesis \eqref{H0} under different circumstances by applying the proposed methods. In all simulations, we evaluate the empirical type-I error and power of test statistics at the significance level $\alpha = 0.05$. The results are calculated with 500 simulation runs. We set sample size $N = 200, 300$ and vary the dimension over $p = 1000, 1500, 2000$ with $p_1 = p_2$. We also consider $\rho=0.3$, $\rho=0.5$, and $\rho=0.7$ to examine the impact of correlation on the tests, representing a weak, moderate, and high correlation, respectively.

\subsection{Study 1. Testing with Lasso and Random Forest}
\label{study1}

To demonstrate the performance of our proposed quadratic-form test statistic $\tilde{T}_n$ described in \eqref{Tn_tilde}, we present three scenarios for the values of $\bbeta$.
\begin{scenario}
\label{scenario1}
We consider $s_{1} = 0$ and $c_1 = 0$ to assess the empirical type-I error.
\end{scenario}

\begin{scenario}
\label{scenario2}
Let $s_{1} = 1, 3, 5$ and $c_1 = 1/s_{1}^{2/3}$ to assess the empirical power with sparse alternatives.
\end{scenario}

\begin{scenario}
\label{scenario3}
We set $s_{1} = \lfloor 30\% p_1\rfloor, \lfloor 50\% p_1 \rfloor, \lfloor 70\% p_1 \rfloor$ and $c_1 = 1/\sqrt{s_{1}}$ (such that $\|\bbeta\|_2 = 1$) to assess the empirical power with dense alternatives.
\end{scenario}
Note that these scenarios with various alternatives are very challenging cases. Actually, Scenario \ref{scenario2} contains extremely sparse nonzero coefficients, and  Scenario \ref{scenario3} includes fairly weak signals. For the setting of the nuisance parameters $\bgamma$ in Models 2 and 3, we take $s_2=20$ and $c_2=0.5$. For the estimation of nuisance functions, we apply the Lasso and the Random Forest (RF). Here the optimal tuning parameter of the Lasso is selected by 10-fold cross-validations using the R-package \texttt{glmnet}, and we exploit the R-package \texttt{randomForest} to implement the RF algorithm.

Empirical rejection rates (ERR) of the proposed test statistic $\tilde{T}_n$ for different models in various settings are summarized in Tables \ref{table_lasso} and \ref{table_rf} (subject to $\rho = 0.5$), where ERR is the proportion of rejected hypotheses among the total 500 Monte Carlo replications. Results based on the settings where $\rho=0.3$ or $\rho=0.7$ are similar and thus moved to the Supplementary Material due to space limitations. See Tables S1 and S2.

We have the following important findings from the simulation results for all scenarios. Firstly, the results of columns ``Size" in Tables \ref{table_lasso} and \ref{table_rf} clearly indicate that the proposed test statistic $\tilde{T}_n$ controls the type-I error very well. Secondly, we observe that $\tilde{T}_n$ suffers from low power against some sparse alternatives as shown in columns ``Power (Sparse)". More specifically, the testing power strongly relies on the sample size, dimension, and number of active elements. For instance, the power is only 0.376 for Model 1 under the settings of $(N, p, s_{1})=(200, 2000, 1)$ in Table \ref{table_lasso}. However, when we have a relatively large sample size of $N=300$, a smaller dimension of $p=1000$, and more active predictors $s_{1} = 5$, the power is increased to 1. Thirdly, $\tilde{T}_n$ has power close to 1 against dense alternatives in linear models. See the results of Model 1 in columns ``Power (Dense)" of Tables \ref{table_lasso} and \ref{table_rf}. It is worth pointing out that $\tilde{T}_n$ still achieves high empirical powers among complex nonlinear structures against dense alternatives. This reflects the merits of tests based on quadratic forms and ML estimation.

\begin{table}[!ht]
\footnotesize
\renewcommand\arraystretch{1}
\centering \tabcolsep 11pt \LTcapwidth 6in
\caption{The empirical reject rates of $\tilde{T}_n$ using Lasso under $\rho=0.5$.}
\label{table_lasso}

\begin{threeparttable}
%\resizebox{\textwidth}{!}{
\begin{tabular}{ccccccccc}
\toprule
&      & Size  & \multicolumn{3}{c}{Power (Sparse)} & \multicolumn{3}{c}{Power (Dense)} \\
\cmidrule(r){3-3} \cmidrule(lr){4-6} \cmidrule(l){7-9}
$N$ & $p$ & $s_{10}$ & $s_{11}$ & $s_{12}$ & $s_{13}$ & $s_{14}$ & $s_{15}$ & $s_{16}$ \\
\midrule
\multicolumn{9}{c}{\bf Model 1}                                                                                   \\
\multirow{3}{*}{200} & 1000 & 0.052 & 0.576 & 0.926 & 0.984              & 1.000 & 1.000 & 1.000              \\
& 1500 & 0.046 & 0.424 & 0.806 & 0.894              & 1.000 & 1.000 & 1.000              \\
& 2000 & 0.054 & 0.376 & 0.712 & 0.854              & 0.998 & 0.994 & 0.996              \\
\cmidrule{2-9}
\multirow{3}{*}{300} & 1000 & 0.060 & 0.902 & 0.996 & 1.000              & 1.000 & 1.000 & 1.000              \\
& 1500 & 0.048 & 0.784 & 0.980 & 0.992              & 1.000 & 1.000 & 1.000              \\
& 2000 & 0.052 & 0.676 & 0.966 & 0.986              & 1.000 & 1.000 & 1.000              \\
\midrule
\multicolumn{9}{c}{\bf Model 2}                                                                                   \\
\multirow{3}{*}{200} & 1000 & 0.046 & 0.474 & 0.820 & 0.916              & 1.000 & 1.000 & 1.000              \\
& 1500 & 0.046 & 0.356 & 0.702 & 0.814              & 1.000 & 0.998 & 0.996              \\
& 2000 & 0.040 & 0.296 & 0.624 & 0.736              & 0.996 & 0.990 & 0.992              \\
\cmidrule{2-9}
\multirow{3}{*}{300} & 1000 & 0.042 & 0.704 & 0.988 & 0.998              & 1.000 & 1.000 & 1.000              \\
& 1500 & 0.040 & 0.586 & 0.946 & 0.984              & 1.000 & 1.000 & 1.000              \\
& 2000 & 0.046 & 0.512 & 0.882 & 0.932              & 1.000 & 1.000 & 1.000              \\
\midrule
\multicolumn{9}{c}{\bf Model 3}                                                                                   \\
\multirow{3}{*}{200} & 1000 & 0.050 & 0.460 & 0.832 & 0.936              & 1.000 & 1.000 & 1.000              \\
& 1500 & 0.050 & 0.374 & 0.714 & 0.812              & 0.998 & 0.998 & 0.998              \\
& 2000 & 0.052 & 0.308 & 0.632 & 0.734              & 0.994 & 0.992 & 0.988              \\
\cmidrule{2-9}
\multirow{3}{*}{300} & 1000 & 0.052 & 0.756 & 0.988 & 1.000              & 1.000 & 1.000 & 1.000              \\
& 1500 & 0.046 & 0.644 & 0.952 & 0.986              & 1.000 & 1.000 & 1.000              \\
& 2000 & 0.058 & 0.520 & 0.898 & 0.958              & 1.000 & 1.000 & 1.000 \\
\bottomrule
\end{tabular}
%}
\begin{tablenotes}[para, flushleft]
\footnotesize
\item \textit{Note.} ``Size" corresponds to the empirical type-I errors. ``Power (Sparse)" and ``Power (Dense)" correspond to the empirical powers with sparse and dense alternatives, respectively. $s_{10}$ represents $s_{1}=0$ (Scenario \ref{scenario1}). $s_{11}$, $s_{12}$ and $s_{13}$ correspond to $s_{1}=1, 3, 5$ (Scenario \ref{scenario2}). $s_{14}$, $s_{15}$ and $s_{16}$ correspond to $s_{1}=\lfloor 30\% p_1\rfloor, \lfloor 50\% p_1\rfloor, \lfloor 70\% p_1\rfloor$ (Scenario \ref{scenario3}).
\end{tablenotes}
\end{threeparttable}
\end{table}

\begin{table}[!ht]
\footnotesize
\renewcommand\arraystretch{1}
\centering \tabcolsep 11pt \LTcapwidth 6in
\caption{The empirical reject rates of $\tilde{T}_n$ using Random Forest under $\rho=0.5$.}
\label{table_rf}

\begin{threeparttable}
%\resizebox{\textwidth}{!}{
\begin{tabular}{ccccccccc}
\toprule
&      & Size  & \multicolumn{3}{c}{Power (Sparse)} & \multicolumn{3}{c}{Power (Dense)} \\
\cmidrule(r){3-3} \cmidrule(lr){4-6} \cmidrule(l){7-9}
$N$ & $p$ & $s_{10}$ & $s_{11}$ & $s_{12}$ & $s_{13}$ & $s_{14}$ & $s_{15}$ & $s_{16}$ \\
\midrule
\multicolumn{9}{c}{\bf Model 1}                                                                                   \\
\multirow{3}{*}{200} & 1000 & 0.048 & 0.550 & 0.890 & 0.968              & 1.000 & 1.000 & 1.000              \\
& 1500 & 0.042 & 0.404 & 0.784 & 0.892              & 1.000 & 0.998 & 1.000              \\
& 2000 & 0.060 & 0.332 & 0.684 & 0.800              & 0.996 & 0.996 & 1.000              \\
\cmidrule{2-9}
\multirow{3}{*}{300} & 1000 & 0.072 & 0.888 & 0.996 & 1.000              & 1.000 & 1.000 & 1.000              \\
& 1500 & 0.074 & 0.716 & 0.978 & 0.994              & 1.000 & 1.000 & 1.000              \\
& 2000 & 0.046 & 0.626 & 0.936 & 0.972              & 1.000 & 1.000 & 1.000              \\
\midrule
\multicolumn{9}{c}{\bf Model 2}                                                                                   \\
\multirow{3}{*}{200} & 1000 & 0.044 & 0.484 & 0.832 & 0.904              & 1.000 & 1.000 & 1.000              \\
& 1500 & 0.054 & 0.348 & 0.692 & 0.806              & 1.000 & 0.996 & 0.994              \\
& 2000 & 0.044 & 0.296 & 0.606 & 0.712              & 0.994 & 0.992 & 0.992              \\
\cmidrule{2-9}
\multirow{3}{*}{300} & 1000 & 0.040 & 0.728 & 0.992 & 0.998              & 1.000 & 1.000 & 1.000              \\
& 1500 & 0.056 & 0.610 & 0.936 & 0.984              & 1.000 & 1.000 & 1.000              \\
& 2000 & 0.050 & 0.524 & 0.860 & 0.934              & 1.000 & 1.000 & 1.000              \\
\midrule
\multicolumn{9}{c}{\bf Model 3}                                                                                   \\
\multirow{3}{*}{200} & 1000 & 0.046 & 0.498 & 0.862 & 0.948              & 1.000 & 1.000 & 1.000              \\
& 1500 & 0.052 & 0.400 & 0.770 & 0.834              & 0.996 & 1.000 & 1.000              \\
& 2000 & 0.054 & 0.316 & 0.676 & 0.780              & 0.996 & 0.988 & 0.994              \\
\cmidrule{2-9}
\multirow{3}{*}{300} & 1000 & 0.056 & 0.818 & 0.992 & 1.000              & 1.000 & 1.000 & 1.000              \\
& 1500 & 0.052 & 0.730 & 0.976 & 0.996              & 1.000 & 1.000 & 1.000              \\
& 2000 & 0.054 & 0.570 & 0.928 & 0.974              & 1.000 & 1.000 & 1.000    \\
\bottomrule
\end{tabular}
%}
\begin{tablenotes}[para, flushleft]
\footnotesize
\item \textit{Note.} ``Size" corresponds to the empirical type-I errors. ``Power (Sparse)" and ``Power (Dense)" correspond to the empirical powers with sparse and dense alternatives, respectively. $s_{10}$ represents $s_{1}=0$ (Scenario \ref{scenario1}); $s_{11}$, $s_{12}$ and $s_{13}$ correspond to $s_{1}=1, 3, 5$ (Scenario \ref{scenario2}); $s_{14}$, $s_{15}$ and $s_{16}$ correspond to $s_{1}=\lfloor 30\% p_1\rfloor, \lfloor 50\% p_1\rfloor, \lfloor 70\% p_1\rfloor$ (Scenario \ref{scenario3}).
\end{tablenotes}
\end{threeparttable}
\end{table}

\subsection{Study 2. Power-enhanced tests}
\label{study2}

In the most challenging case that small $N$, large $p$ and extremely sparse alternatives, our proposed test statistic $\tilde{T}_n$ suffers from low power. To this end, we further investigate the finite sample performance of the power-enhanced testing procedures described in Section \ref{Power-Enhanced Tests}.

\begin{figure}[!ht]
\centering
\includegraphics[width=\textwidth]{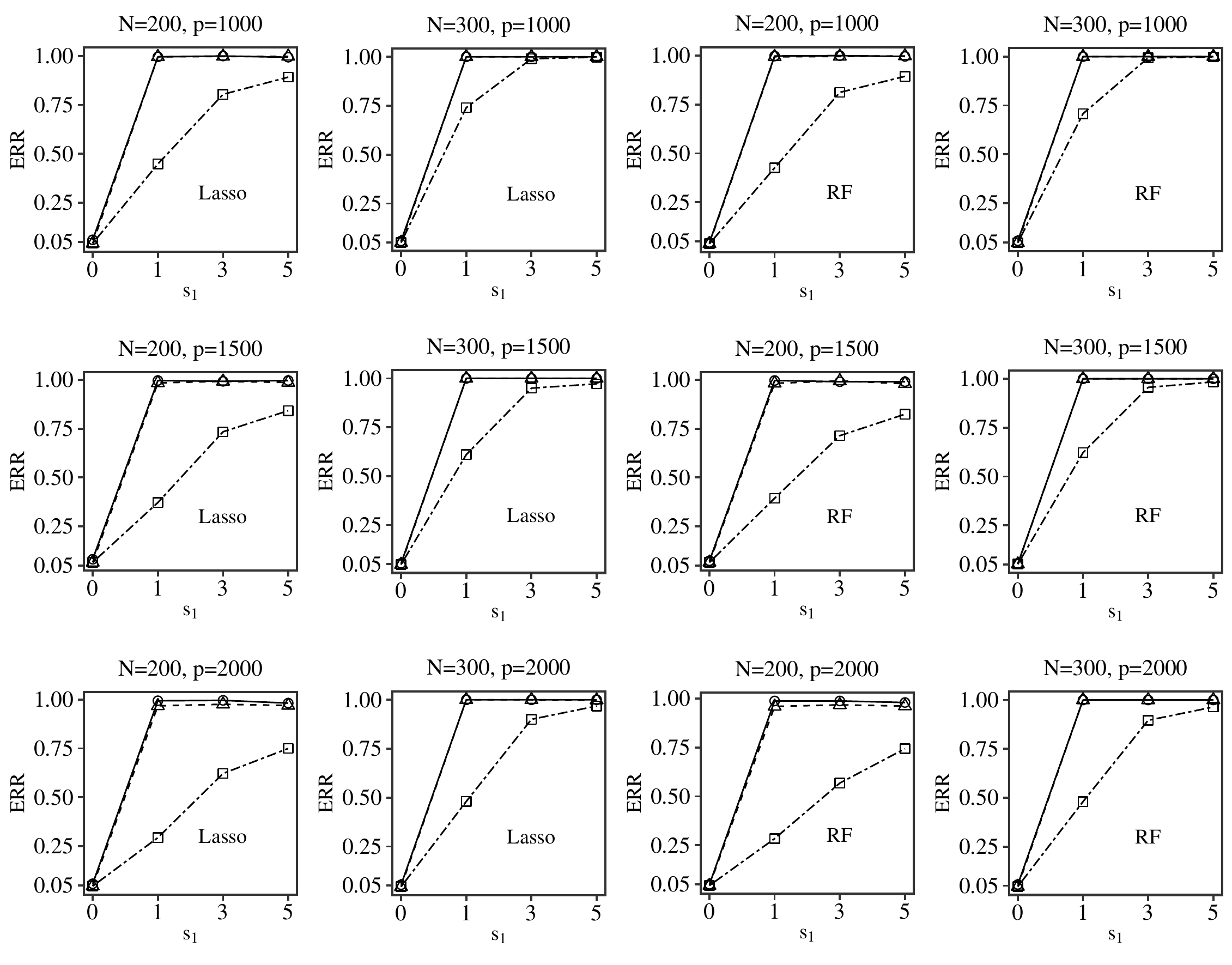}
\caption{The empirical rejection rates of $\tilde{T}_n$ ($\square$ with a two-dashed line), $T_{\mathrm{PE}}^{\mathrm{hard}}$ ($\triangle$ with a dashed line) and $T_{\mathrm{PE}}^{\mathrm{soft}}$ ($\circ$ with a solid line) against the sparsity level $s_{1}$ for {\bf Model 2} under different sample sizes and dimensions. ``Lasso" and ``RF" in figures represent that $\hat{g}(\bZ_i)$ is obtained by Lasso and Random Forest, respectively.}
\label{fig_model2}
\end{figure}

We consider three inference procedures in this part: the quadratic-form test statistic $\tilde{T}_n$ as well as the power-enhanced test statistics $T_{\mathrm{PE}}$ with two thresholds $\delta_{k}^{\mathrm{hard}}$ and $\delta_{k}^{\mathrm{soft}}$, denoted as $T_{\mathrm{PE}}^{\mathrm{hard}}$ and $T_{\mathrm{PE}}^{\mathrm{soft}}$, respectively. Note that we only examine the power enhancement for sparse alternatives. We use the same settings as Scenarios \ref{scenario1} and \ref{scenario2}. Besides, we suggest $\lambda_k = 0.9$ to obtain the hard threshold parameter $\delta_k^{\mathrm{hard}}$ and $(a_{n,p}, R) = (5, 30)$ to estimate the soft threshold parameter $\delta_k^{\mathrm{soft}}$. The comparison results for Model 2 with $\rho=0.5$ are displayed in Figure \ref{fig_model2}. Results for Models 1 and 3 are similar and have been moved to the Supplementary Material. See Figures S1 and S2.

In sum, our proposed power-enhanced test statistics have satisfactory performance. Firstly, $T_{\mathrm{PE}}^{\mathrm{hard}}$ and $T_{\mathrm{PE}}^{\mathrm{soft}}$ do not inflate the empirical size under $\mbH_0$, and still control the type-I error very well, which coincides with the theoretical result in Theorem \ref{thm3.1}. Secondly, the power-enhanced tests $T_{\mathrm{PE}}^{\mathrm{hard}}$ and $T_{\mathrm{PE}}^{\mathrm{soft}}$ substantially promote the testing power of $\tilde{T}_n$, which echoes with the power enhancement property described in Theorem \ref{thm3.2}. Thirdly, the soft threshold $\delta_{k}^{\mathrm{soft}}$ prevails with higher power enhancement compared with the hard threshold $\delta_{k}^{\mathrm{hard}}$ in some cases. Further, the overall patterns for $\rho=0.3$ and $\rho=0.7$ are similar to those in Figure \ref{fig_model2}, and we thereby do not present these results to save space.

\subsection{Study 3. Comparison with the maximum-type statistic}

In this part, we compare our test statistics $\tilde{T}_n$, $T_{\mathrm{PE}}^{\mathrm{hard}}$ and $T_{\mathrm{PE}}^{\mathrm{soft}}$ with the three-step testing procedure based on the studentized statistic in \cite{zhang2017simultaneous}, denoted as $T_{\mathrm{ST}}$. Here, we implement $T_{\mathrm{ST}}$ using R-package \texttt{SILM}, set the number of bootstrap replications as 500 and choose the splitting proportion of 30\% for screening.

The computational expense of $T_{\mathrm{ST}}$ is high since it involves many penalized optimization implementations. Considering the computational limits, we only~study Model 2 and set $(N, p, \rho) = (200, 2000, 0.5)$ under Scenarios \ref{scenario1}-\ref{scenario3}. Additionally, we extend $s_{1} = \{1, 3, 5, 7, 9, 11\}$ in Scenario \ref{scenario2} and $s_{1} = \{200, 300, 400, 500, 600, 700\}$ in Scenario \ref{scenario3}. From the aforementioned simulation results, it can be seen that Lasso and RF have similar performance, we thus just use Lasso in this part.

\begin{figure}[!ht]
\centering
\includegraphics[width=0.81\textwidth]{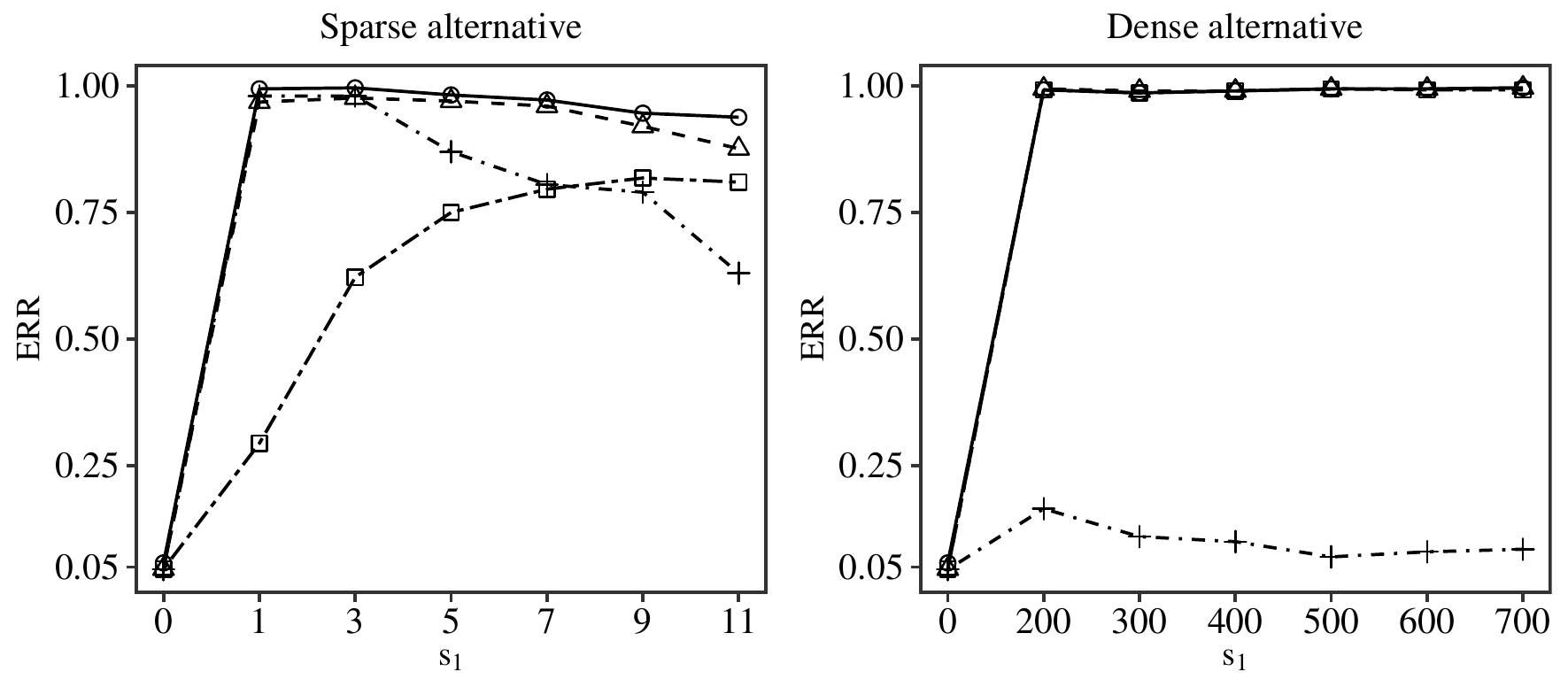}
\caption{The empirical rejection rates of $\tilde{T}_n$ ($\square$ with a two-dashed line), $T_{\mathrm{PE}}^{\mathrm{hard}}$ ($\triangle$ with a dashed line), $T_{\mathrm{PE}}^{\mathrm{soft}}$ ($\circ$ with a solid line) and $T_{\mathrm{ST}}$ ($+$ with a dot-dashed line) against the sparsity level $s_{1}$ for {\bf Model 2} under the sparse (left panel) and dense (right panel) case. Due to the computation limit, the results of $T_{\mathrm{ST}}$ are based on 200 replicates.}
\label{fig4}
\end{figure}

The comparison results for Model 2 is provided in Figure \ref{fig4}, which shows how the empirical reject rates of $\tilde{T}_n$, $T_{\mathrm{PE}}^{\mathrm{hard}}$ and $T_{\mathrm{PE}}^{\mathrm{soft}}$ compare with $T_{\mathrm{ST}}$ at seven values of the sparsity level $s_{1}$. Note that the signal strength $c_1$ decreases with increasing $s_{1}$ since we set $c_1 = 1/s_{1}^{2/3}$ and $c_1 = 1/\sqrt{s_{1}}$  for sparse and dense alternatives, respectively. There are some findings. Firstly, the right panel in Figure \ref{fig4} reveals that our proposed quadratic-form statistic achieves high testing power against dense alternatives. In contrast, the maximum-type method $T_{\mathrm{ST}}$ loses the detection of dense signals. Secondly, $\tilde{T}_n$ lacks the ability to test sparse alternatives, especially $T_{\mathrm{ST}}$ has better performance than $\tilde{T}_n$ when the sparsity level $s_{1}$ is small. However, the test performance of $T_{\mathrm{ST}}$ gradually decreases as the sparsity level $s_{1}$ increases (the signal strength $c_1$ decreases), even inferior to the quadratic-form statistic $\tilde{T}_n$, because of weak signals. Thirdly, it is satisfactory that the proposed power-enhanced tests $T_{\mathrm{PE}}^{\mathrm{hard}}$ and $T_{\mathrm{PE}}^{\mathrm{soft}}$ substantially boost the power of $\tilde{T}_n$ under sparse alternatives and remain high power for dense alternatives. Besides, the proposed power-enhanced tests always have higher power than $T_{\mathrm{ST}}$ in these settings.

Overall, our simulation results demonstrate that the proposed procedures achieve controlled type-I errors and high powers, even when the testing and nuisance covariates are both ultrahigh-dimensional. When the signals are sparse, the power-enhanced test $T_{\mathrm{PE}}$ can greatly improve the performance of the quadratic-form test $\tilde{T}_n$. The power-enhanced test $T_{\mathrm{PE}}$ then provides an innovative and straightforward approach suitable for sparse and dense cases simultaneously. Moreover, ML estimation is an effective tool for handling complex data structures.

\section{Real data analysis}
\label{Real Data Analysis}

We analyze a dataset about riboflavin (vitamin B2) production rate with Bacillus Subtilis, which is a high-dimensional data set made publicly by \cite{buhlmann2014high}. This data set attracts considerable attention and has been systematically analyzed by many authors, for instance, \cite{meinshausen2009p}, \cite{van2014asymptotically}, \cite{javanmard2014confidence} and \cite{yang2022score}. The dataset riboflavin can be obtained from the R-package \texttt{hdi}. It consists of $n = 71$ observations of strains of Bacillus Subtilis. The predictors are the log-expression levels of $p=4088$ genes, and the response is the logarithm of the riboflavin production rate. Same to previous research papers, we standardize the predictors and response to be zero mean and unit variance.

Firstly, we use the sure independence screening procedure based on the distance correlation (DC-SIS, \citeauthor{li2012feature}, \citeyear{li2012feature}) to select genes that possibly impact the response. DC-SIS is implemented here using R-package \texttt{VariableScreening}. The genes selected are denoted as $\bX$, and the other genes are denoted as $\bZ$. Then two natural questions arise: (a) whether the selected genes $\bX$ affect the response $Y$ given the other genes $\bZ$; (b) whether the eliminated genes $\bZ$ by the screening methods are indeed irrelevant to the response $Y$ given the other genes $\bX$. Without loss of generality, we assume that the data set has the same structure as model \eqref{partial_LM}. More specifically, we consider the following regression models,
\begin{align*}
	Y = \bX^{\top}\bbeta_{\bX} + g(\bZ) + \epsilon \quad \text{and} \quad Y = \bZ^{\top}\bbeta_{\bZ} + g(\bX) + \epsilon,
\end{align*}
for questions (a) and (b), respectively. We thereby consider two null hypotheses $\mbH_0: \bbeta_{\bX} = 0$ and $\mbH_0^{\prime}: \bbeta_{\bZ} = 0$ to check whether the chosen genes $\bX$ contribute to the response and whether any significant genes remain in $\bZ$.

Here we pick out the most important 200 genes ($\bX$) by DC-SIS procedure. Since the sample size $n=71$ is small and the result of single data splitting may be unstable, we adopt $\tilde{T}_n$, $T_{\mathrm{PE}}^{\mathrm{hard}}$ and $T_{\mathrm{PE}}^{\mathrm{soft}}$ for testing $\mbH_0$ and $\mbH_0^{\prime}$ with multiple data splitting illustrated in Section S2 in the Supplementary Material. We consider Lasso estimation in our methods. For comparison, we also apply $T_{\mathrm{ST}}$ to make significance tests. Table \ref{table_data1} shows the $p$-values for the considered hypotheses. For the test of $\mbH_0$, our three proposed tests reject the null hypothesis at the level of $\alpha = 0.01$, while $T_{\mathrm{ST}}$ fails to reject it. The results for $\mbH_0^{\prime}$ suggest that the eliminated genes $\bZ$ do not contain any genes that are significantly associated with the response.

\begin{table}[!ht]
\footnotesize
\renewcommand\arraystretch{1}
\centering \tabcolsep 19pt \LTcapwidth 6in
\caption{$p$-values of each test for riboflavin}
\label{table_data1}

\begin{threeparttable}
%\resizebox{\textwidth}{!}{
\begin{tabular}{ccccc}
\toprule
& \multicolumn{4}{c}{Method} \\
\cmidrule(l){2-5}
Test & $\tilde{T}_n$ & $T_{\mathrm{PE}}^{\mathrm{hard}}$ & $T_{\mathrm{PE}}^{\mathrm{soft}}$ & $T_{\mathrm{ST}}$\\
\midrule
$\mbH_0$ & $<$0.001 & $<$0.001 & $<$0.001 & 0.056\\
$\mbH_0^{\prime}$ & 0.330 & 0.330 & 0.330 & 0.174 \\
\bottomrule
\end{tabular}
%}
\begin{tablenotes}[para, flushleft]
\footnotesize
\item \textit{Note.} We adopt the multiple data splitting procedure with $M=30$ times splits.
\end{tablenotes}
\end{threeparttable}
\end{table}

\section{Discussions and conclusions}
\label{Discussions and Conclusions}

In this paper, we develop testing procedures for ultrahigh-dimensional partially linear models. Our procedures allow the testing covariates and nuisance covariates to be both ultrahigh-dimensional and also allow unknown smooth function. To estimate the unknown smooth function, ML methods are used. We construct quadratic-form test statistics by adopting sample-splitting and cross-fitting. We establish the proposed test statistics' asymptotic distributions under the null and local alternative hypotheses. To further improve the performance of proposed test statistics, we introduce power-enhanced tests. Two thresholding rules are introduced for the power-enhanced tests. We prove that the power-enhanced tests can still control the type-I error well, and have larger powers for either sparse or dense alternatives.

In this paper, we model the relationship between $\bX$ and $Y$ linearly while allowing a flexible relationship between $\bZ$ and $Y$. In near future, we aim to allow both flexible modelling of the relationships between $\bX$ and $Y$ and between $\bZ$ and $Y$. New theories and methods are required for this more general model.

\section*{Funding}

The research is supported by National Natural Science Foundation of China (12071038) and Beijing Natural Science Foundation (1212004).

%\newpage
\bibliographystyle{apalike}
\bibliography{bibliography}

\begin{thebibliography}{}

\bibitem[B{\"u}hlmann et~al., 2014]{buhlmann2014high}
B{\"u}hlmann, P., Kalisch, M., and Meier, L. (2014).
\newblock High-dimensional statistics with a view toward applications in
  biology.
\newblock {\em Annual Review of Statistics and Its Application}, 1:255--278.

\bibitem[Cai et~al., 2023]{cai2023tests}
Cai, L., Guo, X., Li, G., and Tan, F. (2023).
\newblock Tests for high-dimensional single-index models.
\newblock {\em Electronic Journal of Statistics}, 17(1):429--463.

\bibitem[Chen et~al., 2023]{chen2022testing}
Chen, J., Li, Q., and Chen, H.~Y. (2023).
\newblock Testing generalized linear models with high-dimensional nuisance
  parameters.
\newblock {\em Biometrika}, 110(1):83--99.

\bibitem[Chen et~al., 2019]{chen2019two}
Chen, S.~X., Li, J., and Zhong, P.-S. (2019).
\newblock Two-sample and {ANOVA} tests for high dimensional means.
\newblock {\em Annals of Statistics}, 47(3):1443--1474.

\bibitem[Chernozhukov et~al., 2018]{chernozhukov2018double}
Chernozhukov, V., Chetverikov, D., Demirer, M., Duflo, E., Hansen, C., Newey,
  W., and Robins, J. (2018).
\newblock Double/debiased machine learning for treatment and structural
  parameters.
\newblock {\em The Econometrics Journal}, 21(1):C1--C68.

\bibitem[Cui et~al., 2018]{cui2018test}
Cui, H., Guo, W., and Zhong, W. (2018).
\newblock Test for high-dimensional regression coefficients using refitted
  cross-validation variance estimation.
\newblock {\em Annals of Statistics}, 46(3):958--988.

\bibitem[Dezeure et~al., 2017]{dezeure2017high}
Dezeure, R., B{\"u}hlmann, P., and Zhang, C.-H. (2017).
\newblock High-dimensional simultaneous inference with the bootstrap.
\newblock {\em Test}, 26(4):685--719.

\bibitem[Du et~al., 2021]{du2021false}
Du, L., Guo, X., Sun, W., and Zou, C. (2021).
\newblock False discovery rate control under general dependence by symmetrized
  data aggregation.
\newblock {\em Journal of the American Statistical Association}, pages 1--15.

\bibitem[Fan et~al., 2012]{fan2012variance}
Fan, J., Guo, S., and Hao, N. (2012).
\newblock Variance estimation using refitted cross-validation in ultrahigh
  dimensional regression.
\newblock {\em Journal of the Royal Statistical Society: Series B (Statistical
  Methodology)}, 74(1):37--65.

\bibitem[Fan et~al., 2020]{fan2020statistical}
Fan, J., Li, R., Zhang, C.-H., and Zou, H. (2020).
\newblock {\em Statistical foundations of data science}.
\newblock Chapman \& Hall.

\bibitem[Fan et~al., 2015]{fan2015power}
Fan, J., Liao, Y., and Yao, J. (2015).
\newblock Power enhancement in high-dimensional cross-sectional tests.
\newblock {\em Econometrica}, 83(4):1497--1541.

\bibitem[G{\"o}tze et~al., 2021]{gotze2021concentration}
G{\"o}tze, F., Sambale, H., and Sinulis, A. (2021).
\newblock Concentration inequalities for polynomials in
  $\alpha$-sub-exponential random variables.
\newblock {\em Electronic Journal of Probability}, 26:1--22.

\bibitem[Guo and Chen, 2016]{guo2016tests}
Guo, B. and Chen, S.~X. (2016).
\newblock Tests for high dimensional generalized linear models.
\newblock {\em Journal of the Royal Statistical Society: Series B (Statistical
  Methodology)}, 78(5):1079--1102.

\bibitem[H{\"a}rdle et~al., 2000]{hardle2000partially}
H{\"a}rdle, W., Liang, H., and Gao, J. (2000).
\newblock {\em Partially linear models}.
\newblock Physica Heidelberg.

\bibitem[Javanmard and Montanari, 2014]{javanmard2014confidence}
Javanmard, A. and Montanari, A. (2014).
\newblock Confidence intervals and hypothesis testing for high-dimensional
  regression.
\newblock {\em Journal of Machine Learning Research}, 15(82):2869--2909.

\bibitem[Li et~al., 2012]{li2012feature}
Li, R., Zhong, W., and Zhu, L. (2012).
\newblock Feature screening via distance correlation learning.
\newblock {\em Journal of the American Statistical Association},
  107(499):1129--1139.

\bibitem[Liu et~al., 2020]{liu2020tests}
Liu, Y., Zhang, S., Ma, S., and Zhang, Q. (2020).
\newblock Tests for regression coefficients in high dimensional partially
  linear models.
\newblock {\em Statistics and Probability Letters}, 163:108772.

\bibitem[Ma et~al., 2021]{ma2021global}
Ma, R., Cai, T., and Li, H. (2021).
\newblock Global and simultaneous hypothesis testing for high-dimensional
  logistic regression models.
\newblock {\em Journal of the American Statistical Association},
  116(534):984--998.

\bibitem[Meinshausen et~al., 2009]{meinshausen2009p}
Meinshausen, N., Meier, L., and B{\"u}hlmann, P. (2009).
\newblock P-values for high-dimensional regression.
\newblock {\em Journal of the American Statistical Association},
  104(488):1671--1681.

\bibitem[Ning and Liu, 2017]{ning2017general}
Ning, Y. and Liu, H. (2017).
\newblock A general theory of hypothesis tests and confidence regions for
  sparse high dimensional models.
\newblock {\em Annals of Statistics}, 45(1):158--195.

\bibitem[Van~de Geer et~al., 2014]{van2014asymptotically}
Van~de Geer, S., B{\"u}hlmann, P., Ritov, Y., and Dezeure, R. (2014).
\newblock On asymptotically optimal confidence regions and tests for
  high-dimensional models.
\newblock {\em Annals of Statistics}, 42(3):1166--1202.

\bibitem[Vansteelandt and Dukes, 2022]{vansteelandt2022assumption}
Vansteelandt, S. and Dukes, O. (2022).
\newblock Assumption-lean inference for generalised linear model parameters.
\newblock {\em Journal of the Royal Statistical Society: Series B (Statistical
  Methodology)}, 84(3):657--685.

\bibitem[Wainwright, 2019]{wainwright2019high}
Wainwright, M.~J. (2019).
\newblock {\em High-dimensional statistics: {A} non-asymptotic viewpoint}.
\newblock Cambridge University Press.

\bibitem[Wang and Cui, 2017]{wang2017generalized}
Wang, S. and Cui, H. (2017).
\newblock Generalized {F}-test for high dimensional regression coefficients of
  partially linear models.
\newblock {\em Journal of Systems Science and Complexity}, 30(5):1206--1226.

\bibitem[Wang and Cui, 2020]{wang2020test}
Wang, S. and Cui, H. (2020).
\newblock Test for high dimensional regression coefficients of partially linear
  models.
\newblock {\em Communications in Statistics - Theory and Methods},
  49(17):4091--4116.

\bibitem[Xu et~al., 2016]{xu2016adaptive}
Xu, G., Lin, L., Wei, P., and Pan, W. (2016).
\newblock An adaptive two-sample test for high-dimensional means.
\newblock {\em Biometrika}, 103(3):609--624.

\bibitem[Yang et~al., 2022]{yang2022score}
Yang, W., Guo, X., and Zhu, L. (2022).
\newblock Score function-based tests for ultrahigh-dimensional linear models.
\newblock {\em arXiv preprint arXiv:2212.08446}.

\bibitem[Yu et~al., 2022]{yu2022power}
Yu, X., Li, D., Xue, L., and Li, R. (2022).
\newblock Power-enhanced simultaneous test of high-dimensional mean vectors and
  covariance matrices with application to gene-set testing.
\newblock {\em Journal of the American Statistical Association}, pages 1--14.

\bibitem[Zhang and Zhang, 2014]{zhang2014confidence}
Zhang, C.-H. and Zhang, S.~S. (2014).
\newblock Confidence intervals for low dimensional parameters in high
  dimensional linear models.
\newblock {\em Journal of the Royal Statistical Society: Series B (Statistical
  Methodology)}, 76(1):217--242.

\bibitem[Zhang and Cheng, 2017]{zhang2017simultaneous}
Zhang, X. and Cheng, G. (2017).
\newblock Simultaneous inference for high-dimensional linear models.
\newblock {\em Journal of the American Statistical Association},
  112(518):757--768.

\bibitem[Zhao et~al., 2023]{zhao2023new}
Zhao, F., Lin, N., and Zhang, B. (2023).
\newblock A new test for high-dimensional regression coefficients in partially
  linear models.
\newblock {\em Canadian Journal of Statistics}, 51(1):5--18.

\bibitem[Zhong and Chen, 2011]{zhong2011tests}
Zhong, P.-S. and Chen, S.~X. (2011).
\newblock Tests for high-dimensional regression coefficients with factorial
  designs.
\newblock {\em Journal of the American Statistical Association},
  106(493):260--274.

\end{thebibliography}

\end{document}